\documentclass[aps,
prl,
10pt,
twocolumn,
superscriptaddress,
longbibliography
]{revtex4-2}

\usepackage{xr}
\usepackage{amsmath,amssymb}
\usepackage{enumitem}
\usepackage{graphicx}
\graphicspath{ {Plots/} }
\usepackage{dcolumn}
\usepackage{bm}
\usepackage{wasysym}
\usepackage{booktabs}

\usepackage[dvipsnames]{xcolor}

\makeatletter
\renewcommand\frontmatter@abstractwidth{\dimexpr\textwidth-1.5in\relax}
\makeatother

\newcommand{\beq}{\begin{equation}}
\newcommand{\eeq}{\end{equation}}
\newcommand{\bea}{\begin{eqnarray}}
\newcommand{\eea}{\end{eqnarray}}

\begin{document}

\title{\LARGE \bf Electronic and chemical phase identification in photoemission experiments using unsupervised machine learning}

\author{Matthew Staab}
	\affiliation{Department of Physics and Astronomy, University of California, Davis, Davis, California $95616$, USA}
\author{Joseph Pandur}
	\affiliation{Department of Physics and Astronomy, University of California, Davis, Davis, California $95616$, USA}
\author{Eli Rotenberg}
	\affiliation{Advanced Light Source, Lawrence Berkeley National Laboratory, Berkeley, CA, USA}
\author{Chris Jozwiak}
	\affiliation{Advanced Light Source, Lawrence Berkeley National Laboratory, Berkeley, CA, USA}
\author{Aaron Bostwick}
	\affiliation{Advanced Light Source, Lawrence Berkeley National Laboratory, Berkeley, CA, USA}
\author{Inna Vishik}
	\affiliation{Department of Physics and Astronomy, University of California, Davis, Davis, California $95616$, USA}
    \affiliation{Materials Sciences Division, Lawrence Berkeley National Lab, Berkeley, CA 94720, USA}
    
\begin{abstract}
\textbf{\begin{center}ABSTRACT\end{center}}
Vacuum ultraviolet photoemission spectroscopies are very information-rich experiments, but due to their surface sensitivity, data are often collected on an initially uncharacterized surface. Traditional raster-grid approaches for locating optimal measurement regions can be time-consuming. In this work, we introduce AARDVARK, a generalizable framework for sample exploration that leverages dimensionality reduction and Gaussian process regression to guide initial sample searches in spatially-resolved photoemission experiments. By utilizing UMAP as a target for a Gaussian process, the algorithm efficiently identifies boundaries of spectroscopically distinct regions, dynamically adapting to variations in sample characteristics. The algorithm enables real-time decision making in measurement selection, optimizes data acquisition, and presents a robust framework for future autonomous sample exploration in photoemission experiments.

\end{abstract}

\date{\today}

\maketitle

\section{Introduction}

\begin{figure*}[htb!]
    \centering
    \includegraphics[width=\linewidth]{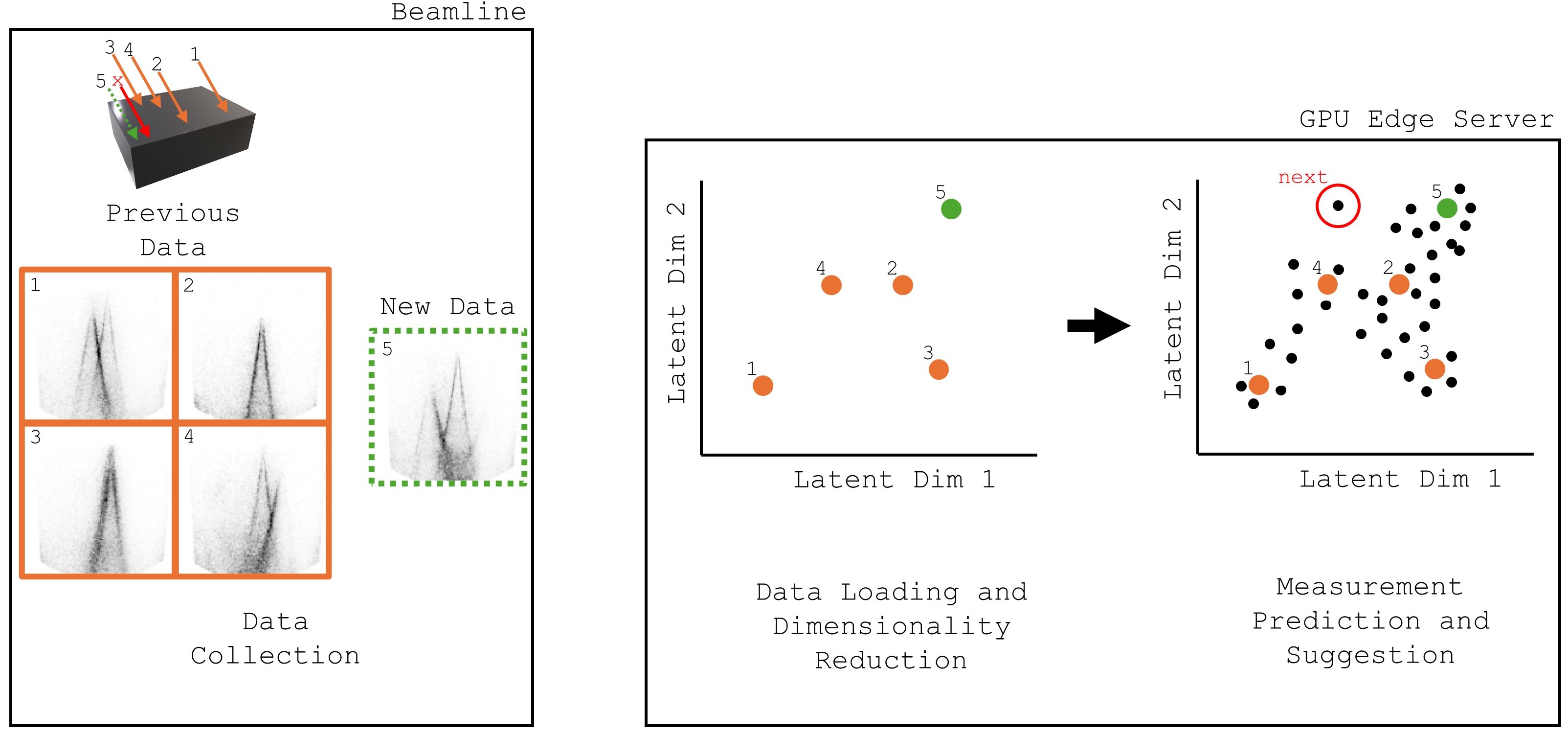}
    \caption[Schematic overview for the AARDVARK system.]{Schematic overview for the AARDVARK system. Data are collected at the beamline end station and saved to disk. The beamline requests a new position to measure and begins measuring a new location. Asynchronously, the GPU edge server is loading all new data and repeating the machine learning pipeline. The machine learning pipeline ends by overwriting the list of next positions to measure with an updated list. We schematically show only one measurement suggestion (red circle) but in practice the server adapts the number of suggestions it creates to ensure there are always positions available for the beamline upon request.}
    \label{fig:overview}
\end{figure*}

Angle-resolved photoemission spectroscopy (ARPES) and x-ray photoelectron spectroscopy (XPS) are techniques based on the photoelectric effect, which elucidate materials' momentum-space electronic structure and chemical structure, respectively ~\cite{ARPESReview,Bagus_2013}.  They can be performed with the same spectrometer on the same specimen, yielding these complementary information streams sequentially. When performed with vacuum ultraviolet (VUV) light, both experiments are extremely surface sensitive, because of the extremely short inelastic mean free path of photoelectrons, typically $1 nm$ or less \cite{Seah_1979,Tanuma_1990}.  This surface sensitivity requires preparation of fresh surfaces prior to each experiment, and even in ultrahigh vacuum ($<5\times10^{-11}$ Torr), surfaces typically degrade or change in 48 hours or less \cite{PhysRevB.81.104521,Biswas_2015}. Thus, the measurement time in any VUV photoemission experiment is inherently limited.

Micro- and nano-focused light sources have enabled measuring mesoscale specimens or targeted regions on a larger specimen, hereby broadening the scope of investigations via high resolution photoemission spectroscopies~\cite{Staab2024, Cattelan2018, Bao2021, Katoch2018, Zhang2018, na2025simulationbasedtrainingframeworkmachinelearning, Rotenberg_2014}. Because the fresh surface is uncharacterized and often heterogeneous, micro-/nano-photoemission experiments typically start with a time-consuming process of scanning the sample surface to categorize the spatial variability of the spectra and find which region is best suited for the science goals of that experiment. The criterion that determines `best' spectral quality is highly variable.  Some examples include: sharpest spectrum, specific local chemical composition/termination, specific thickness or twist angle of an exfoliated thin crystal/heterostructure. The most common technique used to initially characterize the fresh surface is to scan a grid of locations, requiring the experimenter to analyze the data to identify the optimal region. This raster-grid technique is very time-consuming, with most users performing multiple raster-grid scans during a single experiment, with successively smaller measurement area to converge on the ultimate region(s) of interest.
Methods of automating the process of searching for novel information has been a very active research area, including searches of experimental \cite{9492063, 10.1063/5.0172682, doi:10.1021/acs.jpca.3c01685, LIU2023100704, 10.1063/5.0082956, Thomas2022, doi:10.1021/acsnano.0c10239, doi:10.1021/acsnano.1c02104, 10.1063/5.0185362, doi:10.1021/jacs.4c11674, STACH20212702}, material design and synthesis \cite{https://doi.org/10.1002/smtd.202301763, doi:10.1021/acs.accounts.2c00220, Merchant2023, Szymanski2023, 10.1116/6.0004493}, and more general computational \cite{Ziatdinov_2022, KUMAR2025117664, PhysRevB.107.205119, Deringer2021, 10.1063/5.0152367} parameters. All of these efforts aim to reduce the time to discovery by making educated guesses for next steps based on information gathered in the process.
This work describes an alternative to the common raster-grid approach by employing real-time machine learning in the sample search process, compatible with both ARPES and XPS searches. Adaptive Automation and Real Time Data Visualization for All Research Kit (AARDVARK) reduces dimensionality of incoming data for visualization and decision making, and then uses Gaussian process regression to inform the next measurement position, prioritizing more spectrally uncertain spatial regions. By continuously adapting to incoming data, AARDVARK streamlines the early stages of photoemission experiments by more effectively identifying the boundaries between different regions on a heterogeneous sample.

\section{Methods}
\subsection{Overview}
AARDVARK is summarized in Fig.~\ref{fig:overview} and works by launching an asynchronous machine learning loop to constantly provide the best next measurements to the experiment end-station; it was built around the data acquisition architecture at the MAESTRO beamline of the Advanced Light Source. The experiment sends data about the spectrum and location to the server and requests the next measurement location. The AARDVARK server then responds with the next measurement location due to suggestions being sent in batches that are cached for fast response times. At the same time, the AARDVARK server continuously (1) loads newly collected data, (2) processes the data through a machine learning pipeline, (3) evaluates the current state of the sample using a Gaussian process, and (4) computes an acquisition function to suggest a new batch of unmeasured points.  

\subsection{Dataset used in development of AARDVARK}
The data used in this paper were collected on a heterostructure consisting of two single-atomic-layers of graphene stacked on top of each other.  The lower layer is a single crystal spanning the entire sample, and the upper is a continuous polycrystalline layer consisting of finite-sized regions each with a different azimuthal (twist) angle relative to the lower layer. The substrate beneath the heterostructure is SiC.   Original data were recorded on a Scienta R4000 analyzer at the MAESTRO beamline at the Advanced Light Source, using the nano-ARPES branch line.

We performed simulated experiments on a ground truth dataset consisting of 8281 points sampled on a 91$\times$91 grid. At each position of this grid, the following measurements were used in this manuscript: (1) ARPES energy vs momentum cut centered on the Dirac cone of the lower graphene layer, (2) XPS of the Si $2p$ core level. AARDVARK experiments were then simulated on these datasets to produce the results shown in this work. For more information about the ground truth dataset see Ref. \cite{KMeansDrivenGP}.

\subsection{Dimensionality Reduction}
\label{sec:umap_aardvark}

In order to make a decision on which location to measure next, the algorithm must predict what the spectra will look like at a given location. The most naive implementation of this prediction is prohibitively expensive, as it involves fitting a model with 1000 --- 1,000,000 output features depending on the spectra being collected. In order to combat this version of the curse of dimensionality, we use a dimensionality reduction step to compact the information of our spectra into fewer features. In this work, we use Uniform Manifold Approximation and Projection (UMAP)~\cite{UMAP, UMAP_GPU} in order to reduce the feature space that we must train our model on; UMAP has been shown to accurately identify even subtle spectroscopic differences in both ARPES and XPS spectra \cite{sreedhar2025mesoscalevariationschemicalelectronic}. The implementation of UMAP used here is from the RAPIDS cuML package~\cite{RAPIDS}, which provides a GPU accelerated implementation of the algorithm from refs~\cite{UMAP, UMAP_GPU}. With this acceleration, we are able to fit and transform hundreds of full-resolution (approximately 1000$\times$1000) ARPES spectra in roughly one second. The rapidity of prediction using GPU acceleration allows for our algorithm to provide new optimal measurement suggestions every few seconds, depending on how many spectra are being collected and their size.

\begin{figure}[htb!]
    \centering
    \includegraphics[width=\linewidth]{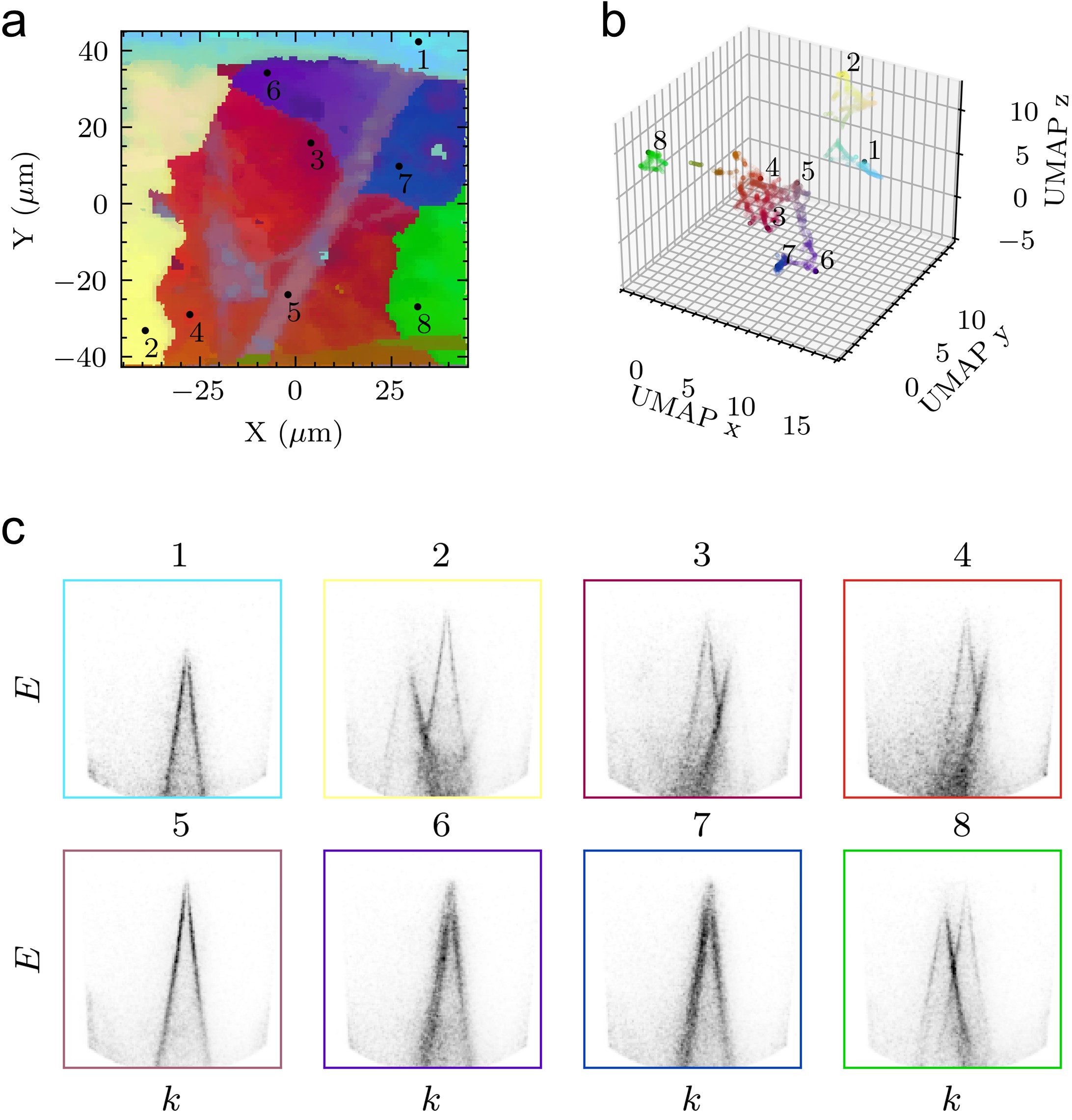}
    \caption[Colorized UMAP embeddings.]{Colorized UMAP embeddings of ground truth dataset. (a) Spatial plot of the UMAP embedded ARPES data. The colors are made by scaling the range of the UMAP x, y, z coordinates to [0, 255] such that they can be displayed in the red, green, and blue channels of an image respectively. (b) The set of all ARPES spectra shown in their UMAP dimensionality reduced space with the markers colored based on the same procedure described in (a). (c) ARPES spectra taken at the locations indicated by the black marker and numbers in (a) and (b). }
    \label{fig:umap-color}
\end{figure}

The choice of embedding into a 3-dimensional UMAP space enables immediate user feedback on the progress of the experiment by visualizing the 3 UMAP dimensions as red, green, and blue channels of an image as shown in Figure~\ref{fig:umap-color}(a). For the visualization, each of the 3 UMAP dimensions (x, y, z) are scaled to the same range [0, 255] such that they can be visualized as a color. The same embedding can also be visualized as points in 3D space, as shown in Fig. \ref{fig:umap-color}.

With the UMAP visualization, similar colors generally have similar spectra, while distinct colors generally have distinct spectra. Because the UMAP algorithm is non‐deterministic and heavily dependent on both the input spectra and the random seed, each run of the UMAP visualization will yield different color assignments for the regions, and it is not possible to set a consistent color scale during the evolution of a scan.  The nonlinearity of the UMAP embedding favors preserving small differences in the data, such that colors which are very similar represent spectra which are similar.  In contrast, one cannot interpret large differences in color as large differences in spectra.

In Figure~\ref{fig:umap-color} (c), we show ARPES spectra from the ground-truth data sets, demonstrating how different colors of the UMAP embedding correspond to different apparent spectra.  The SiC substrate is an insulator with the valence band maximum much deeper than the graphene Dirac cones being probed. The single-crystal graphene layer has a single orientation over the whole field of view, which is spectroscopically seen as the persistent Dirac cone that is near the center of the momentum window. The top polycrystalline graphene layer provides a secondary Dirac cone at lower binding energy whose momentum position varies with the twist angle between the two sheets of graphene. The twist angle and bonding strength of the two layers also affects the Fermi level, which varies the binding energy of both Dirac cones.  Various regions in the UMAP color visualization (Fig. \ref{fig:umap-color}(a)) correspond to the following local characteristics in (c): (1) monolayer spectra, (2, 3, 4, 8) various large twist angles between the polycrystalline top sheet and single crystal bottom sheet, (5) regions where the top layer has cracked and reveals only the bottom layer, and (6, 7) regions with very small twist angles.

\subsection{Gaussian Process Regression}
\begin{figure*}[htb!]
    \centering
    \includegraphics[width=\linewidth]{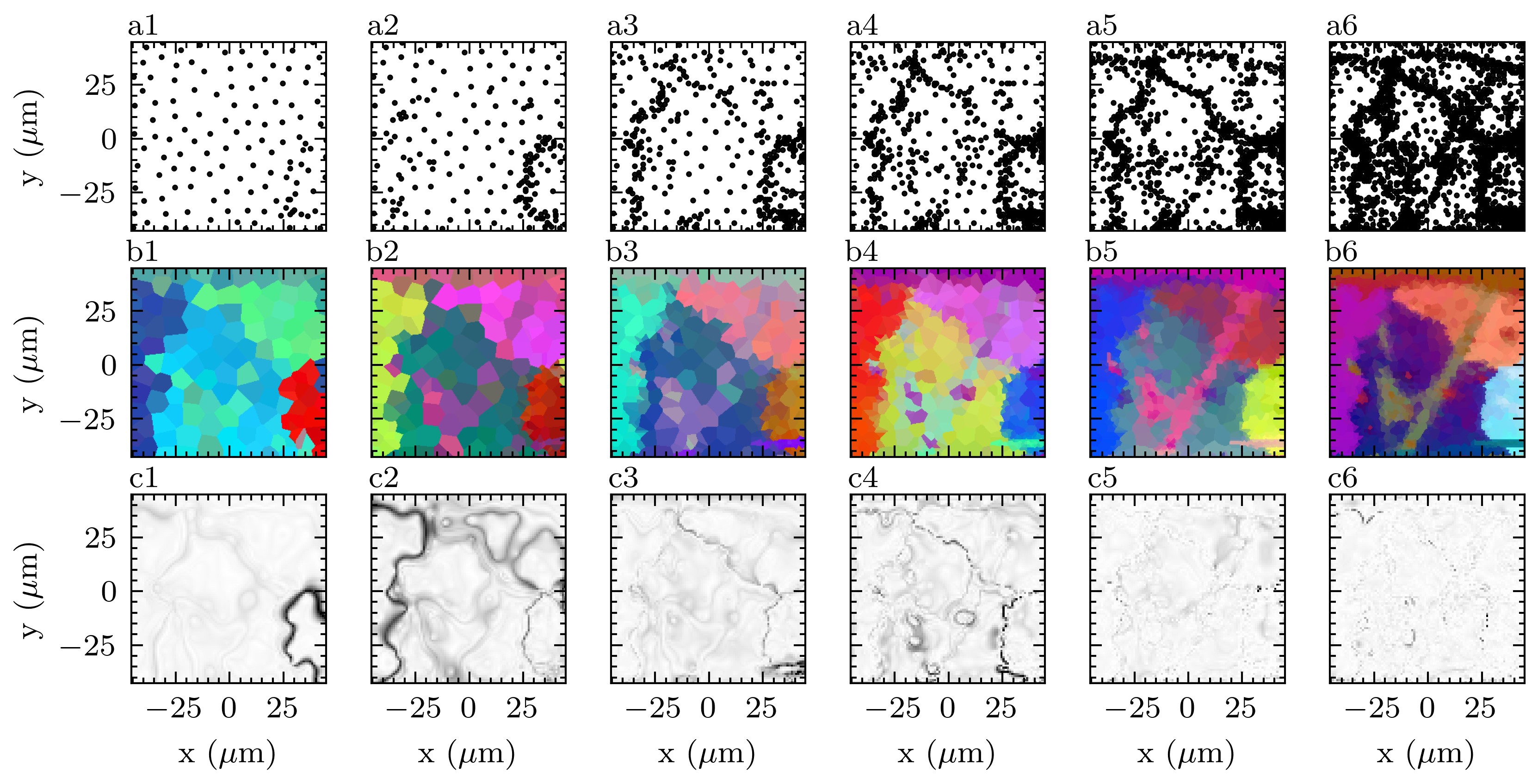}
    \caption[Progression of an AARDVARK scan on ARPES data.]{Progression of an AARDVARK scan on ARPES data. (a) Coordinates of sampled data with (125, 200, 400, 600, 1000, 2000) points measured are shown in (a1-a6) respectively. (b) Nearest-neighbor colorized 3-dimensional UMAP embeddings of ARPES spectra collected at the points shown in (a). Different colors indicate different spectral features as found by the dimensionality reduction method UMAP. (c) The acquisition function that is used to determine the priority of following measurements. Darker pixels indicate a higher measurement priority.}
    \label{fig:arpes}
\end{figure*}

Gaussian process regression is a non‐linear fitting method that excels when the dataset is small and the underlying functional form is unknown. The process of spatially exploring ARPES or XPS spectra of a sample fits both of these criteria because we start from no initial data and do not know the sample's spatial makeup prior to the experiment. Here, we utilize GPyTorch, a GPU accelerated implementation of Gaussian processes from Ref.~\cite{GPYTORCH}. Many in-depth explanations of the operation of a Gaussian process exist, with some even targeting ARPES specifically, so curious readers are encouraged to explore those resources~\cite{Wang2023, Rasmussen2005}. For this work, we use the following definition of a Gaussian process: a model that can perform a non-linear fit to data and provide expected values and variances of measurements at inputs that have not been measured. While neural networks are another popular method for non‐linear fitting, they are less suited to our use case because they require a large amount of training data relative to the number of fitting parameters. Due to the lack of data inherent in searching a new sample, a Gaussian process is better suited to the task of exploration than a neural network.

Identifying the interesting regions of a sample can be performed by using a multitask Gaussian process~\cite{MULTITASK_GP} hand-tuned to your experiment. For ARPES spatial scans, these hand-tuned tasks may be related to the intensity and sharpness of the image. For XPS spatial scans, the tasks may be more specific, involving the lineshapes or binding energies of the resolved peaks, or simply their intensity. In this work, we use UMAP for dimensionality reduction on both ARPES and XPS spectra, applying the resulting UMAP coordinates as targets for the Gaussian process. The use of the UMAP coordinates as a Gaussian process target allows the algorithm to be independent of the particulars of a given experiment or spectrum. To emphasize exploration of unknown spectra, the algorithm fits a multitask Gaussian process to the UMAP coordinates and prioritizes the next measurements based on those whose predicted UMAP coordinates are furthest from the coordinates of any existing measurement. This process favors exploration of borders first by definition because the Gaussian process fits a continuous function to the existing measurements. Once the borders have been sampled up to the user's defined resolution, the algorithm is forced to explore the interior of the regions because duplicated measurements are not allowed within a tolerance specified by the user. The acquisition can be described by the following equation.

\begin{equation}
    \label{acquisition-function}
    f_a(\mathbf{x}) = \min( \lVert \textbf{U}(\textbf{x}) - \textbf{U}(\textbf{x}_\text{obs}) \rVert )
\end{equation}

Where $\mathbf{U}(\mathbf{x})$ is the UMAP embedding predicted by the Gaussian process for the position $\mathbf{x}$ and $\mathbf{U}(\mathbf{x}_\text{obs})$ are the UMAP embeddings of the current measurements that the Gaussian process was trained on.

Further details of the implementation of AARDVARK can be found in Supplementary Materials.


\section{Results}

\begin{figure*}[htb!]
    \centering
    \includegraphics[width=\linewidth]{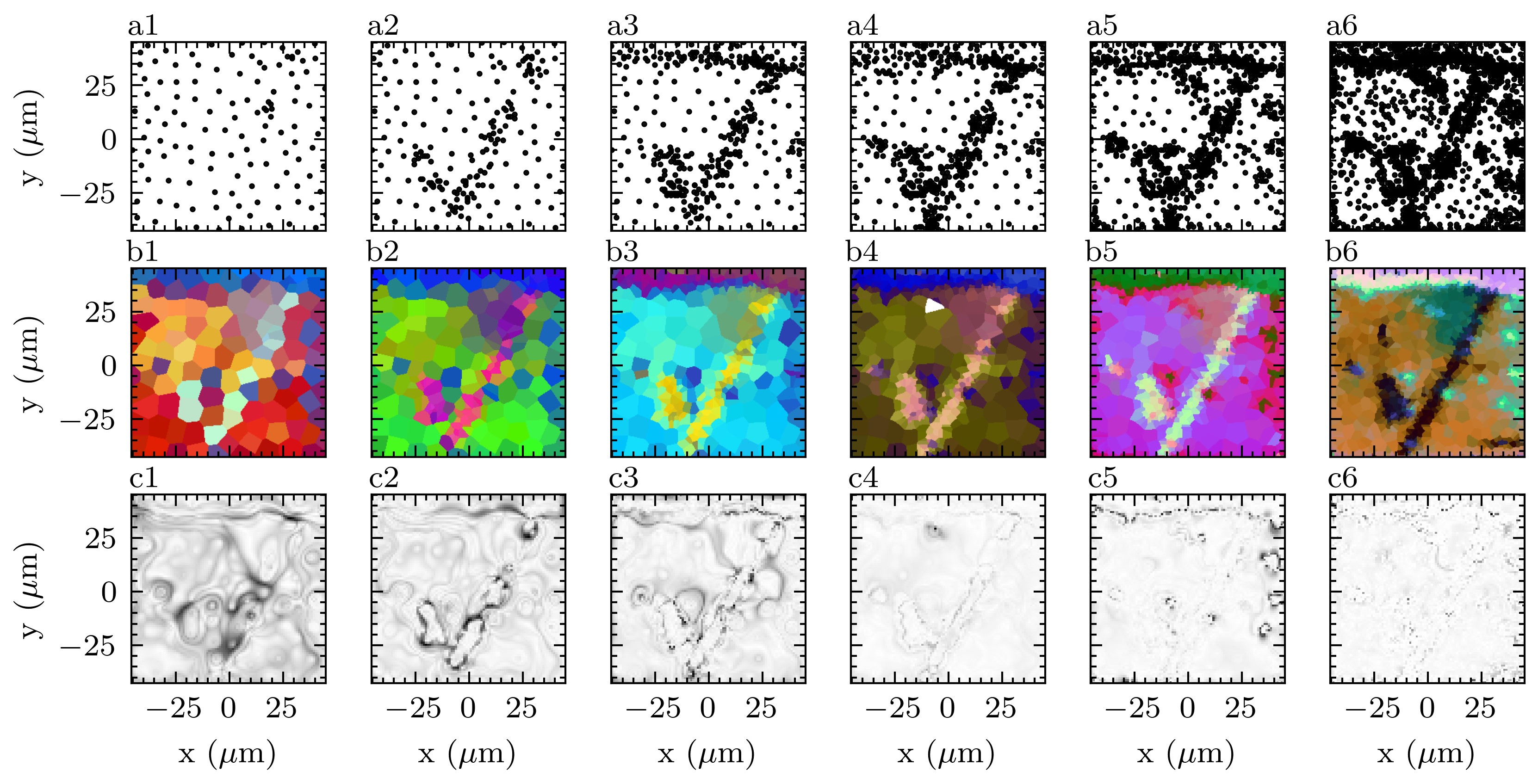}
    \caption[Progression of an AARDVARK scan on XPS data.]{Progression of an AARDVARK scan on XPS data. (a) Coordinates of sampled data with (125, 200, 400, 600, 1000, 2000) points measured are shown in (a1-a6) respectively. (b) Nearest-neighbor colorized 3-dimensional UMAP embeddings of XPS spectra of the Si $2p$ core level collected at the points shown in (a). Different colors indicate different spectral features as found by the dimensionality reduction method UMAP. (c) The acquisition function that is used to determine the priority of following measurements. Darker pixels indicate a higher measurement priority.}
    \label{fig:xps}
\end{figure*}

In Figure~\ref{fig:arpes}, we show the progression of an AARDVARK experiment set to collect the ARPES spectra of the sample. The top row of panels (a1-a6) shows the locations of measurements after (125, 200, 400, 600, 1000, 2000) measurements were taken. AARDVARK made new decisions for measurements every six seconds on average over the course of this experiment. The middle row of panels (b1-b6) shows the colorized UMAP embedding of the ARPES spectra collected at the positions shown in panels (a1-a6). The lower row (c1-c6) shows the acquisition function described in equation \ref{acquisition-function}, where the darker regions represent the highest acquisition value and highest priority for the next measurement.

There are six dominant regions observed in the ARPES experiment spectra that are shown sampling the full dataset in Figure~\ref{fig:umap-color} with their characteristic spectra shown in panel (c): (1, 2, 3/4, 5, 6/7, 8). Grouped regions (3/4, 6/7) show qualitative spectral similarity and similar UMAP colorization. After 400 measurements, AARDVARK is able to find and define the borders of these major regions, though subsequent measurements help define the boundaries with greater precision.

In Figure~\ref{fig:xps}, we show a simulated experiment on the XPS spectra of the Si-$2p$ core level, originating from the substrate, measured on the same system. Panels (a1-a6) again show the positions of the collected measurements after (125, 200, 400, 600, 1000, 2000) points had been taken. Figure \ref{fig:xps} (b1-b6) shows the colorized UMAP embedding of the XPS spectra collected at the points shown in (a1-a6). Panels (c1-c6) show the evolution of the acquisition function over the course of the simulated XPS experiment.

As indicated by their UMAP colorization in Figure~\ref{fig:xps} (b1-b6), the region surrounding the check-mark-shaped feature is observed to be more uniform in its Si $2p$ XPS spectrum than that seen in the ARPES spatial map in Figure~\ref{fig:arpes}. In the Si $2p$ XPS spectra, we observe only four dominant regions---pink, green, brown, and black in (b6)---along with some small scattered features in teal. The single central Dirac cone characterizing the checkmark feature (Figure~\ref{fig:umap-color} (c5)) suggests that in this region the polycrystalline top layer of graphene is absent and the monocrystalline graphene is exposed. This difference in film thickness may explain why the check-mark-shaped feature is the primary feature contrasted from its surroundings in the XPS colorized UMAP of Figure~\ref{fig:xps}.  While AARDVARK is designed to work robustly with both ARPES and XPS spectra, the region boundaries it identifies might be different in one channel or another, depending on the physical origin of the signal.

\begin{figure*}[htb!]
    \centering
    \includegraphics[width=\linewidth]{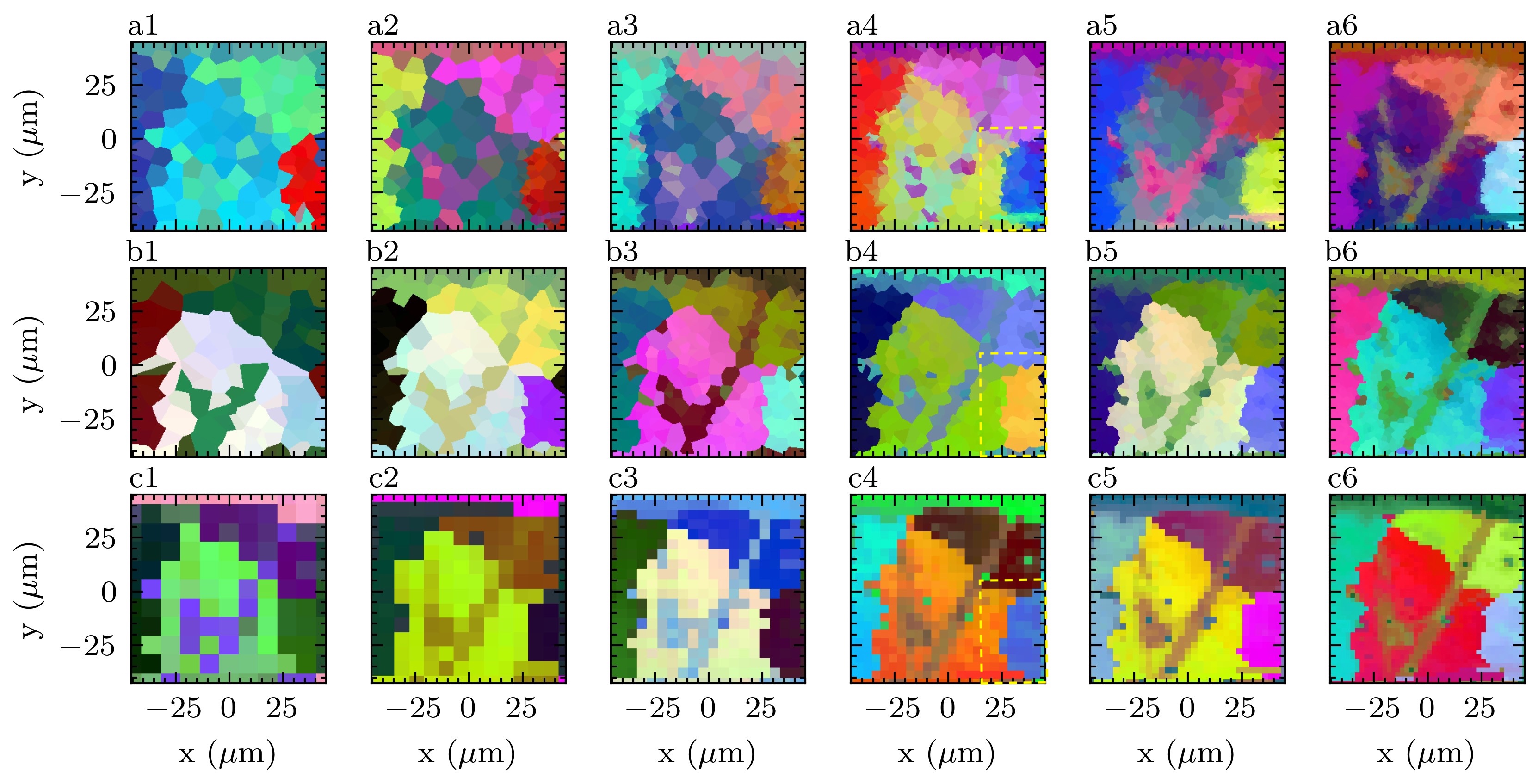}
    \caption[Comparison of AARDVARK operation on ARPES data with random or grid search.]{Comparison of AARDVARK operation on ARPES data with random or grid search. Rows (a-c) show the progression of scans using AARDVARK, random positions, and grid searches respectively. Panels (1–6) display colorized UMAP visualizations for scans (AARDVARK, random, and grid) at approximately 125, 200, 400, 600, 1000, and 2000 measurement points, respectively. In the case of a grid search the closest square number is used to simulate a scan of approximately the same number of points (e.g. 121 instead of 125). Dotted yellow lines on the fourth column represent the boundaries of the cropped UMAP embeddings analyzed in Figure \ref{fig:boundaries_comparison}}.
    \label{fig:comparison}
\end{figure*}

\section{Discussion}
Autonomous, ML/AI‐powered, and human‐in‐the‐loop control systems are a rapidly advancing research topic that is accelerating experiments across fields. Previous post‐experiment analyses have shown that K‐means classification and Gaussian processes can effectively explore a sample using ARPES~\cite{KMeansDrivenGP}. Additionally, autonomous ARPES has been performed by employing a Gaussian process on tasks hand-tuned to frequent goals of high intensity and sharp spectra~\cite{Agustsson2024}. Other works have demonstrated that real time ML can dramatically speed up performance in X-ray microscopy experiments at synchrotrons~\cite{AIDrivenFAST}. AARDVARK deploys a modular architecture that allows for extension to new ML frameworks and demonstrates that UMAP as a general purpose Gaussian process target provides fast exploration of samples for both ARPES and XPS.

The most common method of sample exploration is a raster-grid search where a user specifies a range and step size to explore the sample in a grid. While common the raster-grid search requires the user to know the exact region to explore and at what resolution before performing the search. If too low of a resolution is chosen users will have to take another search over an updated range with better resolution. If too high of a resolution is chosen then the user must wait a long time to get information about the sample because the search sweeps across the sample.

An alternative method which is rarely used is a random search of the sample; in the context of this manuscript, we include the random search as a contrast to AARDVARK, because it samples a distribution of points across the sample absent of learning-based approaches. A random search provides rapid feedback across the entire sample surface but fails to focus on important features due to its indiscriminate nature. A random search can also waste time by repeating measurements nearby other measurements in regions that can be expected to be relatively homogeneous from inspection of the prior data. Another disadvantage of a random search is that it requires large motions of the motors. These large motions can take longer or be less precise in some cases. In future implementations, AARDVARK could weight its point selection order by motor travel distance, reducing long motor motions.

In Figure~\ref{fig:comparison} we show a comparison of an AARDVARK scan (a1-a6), a random search (b1-b6) and a set of grid searches (c1-c6) which all contain roughly the same number of measurements per column (125, 200, 400, 600, 1000, 2000) respectively. After collecting 400 --- 600 points, the AARDVARK scan (a3–a4) exhibits as much or more spatial information about the sample surface compared to the random or grid scans at 1000 --– 2000 points (b6, c6). It should be noted as well that in a real world test a grid search would not appear as it does in the lower row (c1-c6) because the user must first specify the total number of points desired. The more direct comparison of a grid scan would then be a scan with the resolution of perhaps (c5) which only has seen 60\% of the sample when compared to AARDVARK in (a4). AARDVARK appears to outperform the other techniques in locating and defining the boundaries of the regions as well as finding small features like the crack on the lower right side of the sample. Accurately identifying boundaries is an efficient way of selecting target measurement positions inside the boundary, but it can have scientific merit on its own, as 2D quantum phenomena often have defining 1D boundary states  \cite{Konig:QSHI2008,Tokura_2019}. The spatial extent of these boundary states is typically $<1 nm$, much smaller than even the state-of-the-art nanoARPES spot size ($\approx50 nm$), but accurate identification of the boundary could provide enough statistics to observe momentum space signatures of these boundary states via their cumulative different from interior spectra.

\begin{figure}
    \centering
    \includegraphics[width=\linewidth]{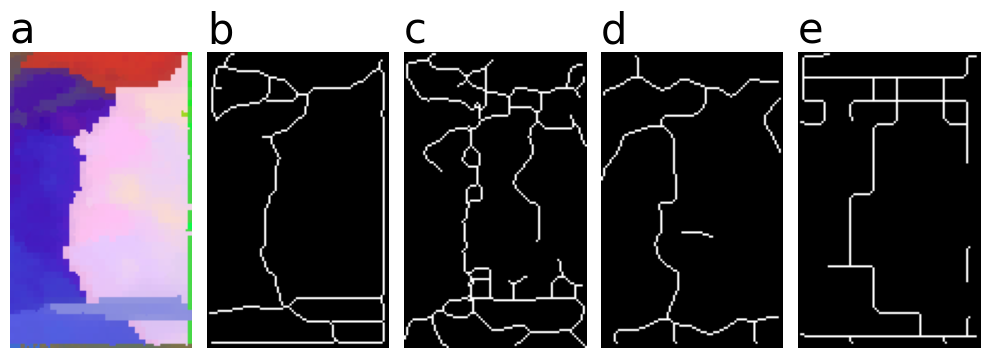}
    \caption{Examining boundaries in cropped UMAP embeddings. From left to right: (a) portion of ground truth data indicated by yellow box in Fig. \ref{fig:comparison} column 4 (b) ground truth boundaries (c) AARDVARK boundaries (d) random positions boundaries, and (e) grid search boundaries. For the latter three, the 600 points trial is used. White meander denotes boundary.}
    \label{fig:boundaries_comparison}
\end{figure}

The boundaries determined from the three search techniques as well as ground truth are summarized in Fig. \ref{fig:boundaries_comparison}, and the determination of the boundary positions is discussed in supplementary materials.  For clarity, we focus on the lower right portion of the ground truth image.  The boundary locations relative to ground truth are examined using the following metrics: precision, which we define as the fraction of predicted boundary pixels that align with the ground truth boundary pixels, and recall, which we define as the fraction of ground truth pixels that are detected by the predicted boundary lines. When computing each of these quantities, a margin of error measured by distance separation between the ground truth and predictions is allowed.  Within these two metrics there are false positives (FP) and true positives (TP), corresponding to correct and incorrect determination of a boundary by both metrics (see supplementary materials).


\begin{table}[htbp]
    \centering
    \caption{Precision (Prec) and recall (Rec) values with error margin of 1 $\mu$m.}
    \label{tab:precision_recall_3}
    \begin{tabular}{l c c c c c c}
        \toprule
        Method & Prec & TP (Pre) & FP & Rec & TP (Rec) & FN \\
        \midrule
        AARDVARK & 0.49739 & 382 & 386 & 0.64545 & 355 & 195 \\
        Random   & 0.43811 & 177 & 227 & 0.37090 & 204 & 346 \\
        Grid     & 0.36991 & 209 & 356 & 0.29272 & 161 & 389 \\
        \bottomrule
    \end{tabular}
\end{table}

Precision and recall for all three methods are summarized in Table \ref{tab:precision_recall_3}, using a margin of error of of $1 \mu m$.  Because the goal is to accurately detect ground truth boundaries, recall is a better quantity of interest, for which AARDVARK significantly outperforms the other two methods.


Figures \ref{fig:arpes} and \ref{fig:xps} also highlight the general purpose utility of AARDVARK as the same algorithm with no changes is able to effectively search samples with diverse spectroscopic signatures. In the XPS spectra of the Si $2p$ core level, spatial inhomogeneities originate from varied thickness of the heterostructure but this metric is not sensitive to changes in twist angle. In contrast, the ARPES spectra show large spatial inhomogeneity where differences in the twist angle changes are visualized as different location of the polycrystalline graphene Dirac cone. The generality of the UMAP embedding approach allows for AARDVARK to account for this difference and identify the different distinguishing structures of the ARPES and XPS data.  This difference can be seen in the locations of the measurements in Figure~\ref{fig:arpes} (a1-a6) and Figure~\ref{fig:xps} (a1-a6). A raster-grid search or a random search would treat both experiments and datasets in the same manner. In addition to potentially saving user beamtime during an experiment the AARDVARK scan can be used when there is a concern over beam radiation damage because it gives the user spatially resolved spectroscopic information with fewer measurements.

\section{Conclusions}
Here we presented AARDVARK, a generalizable framework for sample exploration that leverages dimensionality reduction and Gaussian process regression to guide initial sample searches in ARPES and XPS experiments. By utilizing UMAP as a target for a Gaussian process, AARDVARK identifies boundaries of spectroscopically distinct regions, dynamically adapting to variations in sample characteristics for both ARPES and XPS measurements. This approach enables a more efficient understanding of a sample’s spatial structure compared to traditional raster‐grid strategies. The key advantages of this approach are: potentially limiting radiation damage during identification process, ability to leverage unique information channels provided by ARPES and XPS, and more efficient identification of boundary regions.



\section{Acknowledgments}
This work was supported by  the US-Israel Binational Science Foundation Grant 2020067 and by the Gordon and Betty Moore Foundation, grant DOI 10.37807/GBMF12957. This research used resources of the Advanced Light Source, a U.S. DOE Office of Science User Facility under contract no. DE-AC02-05CH11231.  MS was supported in part by an ALS Doctoral Fellowship in Residence.

\bibliography{bibliography}

@article{ARPESReview,
  title = {Angle-resolved photoemission studies of quantum materials},
  author = {Sobota, Jonathan A. and He, Yu and Shen, Zhi-Xun},
  journal = {Rev. Mod. Phys.},
  volume = {93},
  issue = {2},
  pages = {025006},
  numpages = {72},
  year = {2021},
  month = {May},
  publisher = {American Physical Society},
  doi = {10.1103/RevModPhys.93.025006},
  url = {https://link.aps.org/doi/10.1103/RevModPhys.93.025006}
}

@article{UMAP,
  title={{UMAP}: Uniform manifold approximation and projection for dimension reduction},
  author={McInnes, Leland and Healy, John and Melville, James},
  journal={arXiv preprint arXiv:1802.03426},
  year={2018}
}

@inproceedings{UMAP_GPU,
  title={Bringing {UMAP} closer to the speed of light with GPU acceleration},
  author={Nolet, Corey J and Lafargue, Victor and Raff, Edward and Nanditale, Thejaswi and Oates, Tim and Zedlewski, John and Patterson, Joshua},
  booktitle={Proceedings of the AAAI Conference on Artificial Intelligence},
  volume={35},
  pages={418--426},
  year={2021}
}

@Manual{RAPIDS,
              title = {RAPIDS: Libraries for End to End GPU Data Science},
              author = {RAPIDS Development Team},
              year = {2023},
              url = {https://rapids.ai}
            }

@article{GPYTORCH,
  title={Gpytorch: Blackbox matrix-matrix gaussian process inference with gpu acceleration},
  author={Gardner, Jacob and Pleiss, Geoff and Weinberger, Kilian Q and Bindel, David and Wilson, Andrew G},
  journal={Advances in neural information processing systems},
  volume={31},
  year={2018}
}

@article{MULTITASK_GP,
  title={Multi-task Gaussian process prediction},
  author={Bonilla, Edwin V and Chai, Kian and Williams, Christopher},
  journal={Advances in neural information processing systems},
  volume={20},
  year={2007}
}

@article{KMeansDrivenGP,
  author  = {{Charles N Melton, Marcus M Noack, Taisuke Ohta, Thomas E Beechem, Jeremy Robinson, Xiaotian Zhang, Aaron Bostwick, Chris Jozwiak, Roland J Koch, Petrus H Zwart, Alexander Hexemer and Eli Rotenberg}},
  title   = {K-means-driven Gaussian Process data collection for angle-resolved photoemission spectroscopy},
  journal = {Machine Learning: Science and Technology},
  year    = {2020},
  volume  = {1},
  number  = {4},
  url     = {https://iopscience.iop.org/article/10.1088/2632-2153/abab61/meta}
}

@article{AIDrivenFAST,
	author = {Kandel, Saugat and Zhou, Tao and Babu, Anakha V. and Di, Zichao and Li, Xinxin and Ma, Xuedan and Holt, Martin and Miceli, Antonino and Phatak, Charudatta and Cherukara, Mathew J.},
	da = {2023/09/07},
	doi = {10.1038/s41467-023-40339-1},
	id = {Kandel2023},
	isbn = {2041-1723},
	journal = {Nature Communications},
	number = {1},
	pages = {5501},
	title = {Demonstration of an AI-driven workflow for autonomous high-resolution scanning microscopy},
	ty = {JOUR},
	url = {https://doi.org/10.1038/s41467-023-40339-1},
	volume = {14},
	year = {2023}}

@ARTICLE{Wang2023,
  author={Wang, Jie},
  journal={Computing in Science and Engineering}, 
  title={An Intuitive Tutorial to Gaussian Process Regression}, 
  year={2023},
  volume={25},
  number={4},
  pages={4-11},
  doi={10.1109/MCSE.2023.3342149}}

@book{Rasmussen2005,
    author = {Rasmussen, Carl Edward and Williams, Christopher K. I.},
    title = {Gaussian Processes for Machine Learning},
    publisher = {The MIT Press},
    year = {2005},
    month = {11},
    isbn = {9780262256834},
    doi = {10.7551/mitpress/3206.001.0001},
    url = {https://doi.org/10.7551/mitpress/3206.001.0001},
}

@article{Agustsson2024,
    author = {Agustsson, Steinn Ymir and Jones, Alfred J. H. and Curcio, Davide and Ulstrup, Soren and Miwa, Jill and Mottin, Davide and Karras, Panagiotis and Hofmann, Philip},
    title = {Autonomous micro-focus angle-resolved photoemission spectroscopy},
    journal = {Review of Scientific Instruments},
    volume = {95},
    number = {5},
    pages = {055106},
    year = {2024},
    month = {05},
    issn = {0034-6748},
    doi = {10.1063/5.0204663},
    url = {https://doi.org/10.1063/5.0204663}
}

@Article{Cattelan2018,
AUTHOR = {Cattelan, Mattia and Fox, Neil A.},
TITLE = {A Perspective on the Application of Spatially Resolved ARPES for 2D Materials},
JOURNAL = {Nanomaterials},
VOLUME = {8},
YEAR = {2018},
NUMBER = {5},
ARTICLE-NUMBER = {284},
URL = {https://www.mdpi.com/2079-4991/8/5/284},
PubMedID = {29702567},
ISSN = {2079-4991},
DOI = {10.3390/nano8050284}
}

@Article{Bao2021,
author={Bao, Changhua
and Zhang, Hongyun
and Li, Qian
and Zhou, Shaohua
and Zhang, Haoxiong
and Deng, Ke
and Zhang, Kenan
and Luo, Laipeng
and Yao, Wei
and Chen, Chaoyu
and Avila, Jos{\'e}
and Asensio, Maria C.
and Wu, Yang
and Zhou, Shuyun},
title={Spatially-resolved electronic structure of stripe domains in IrTe2 through electronic structure microscopy},
journal={Communications Physics},
year={2021},
month={Oct},
day={19},
volume={4},
number={1},
pages={229},
issn={2399-3650},
doi={10.1038/s42005-021-00733-x},
url={https://doi.org/10.1038/s42005-021-00733-x}
}

@Article{Katoch2018,
author={Katoch, Jyoti
and Ulstrup, S{\o}ren
and Koch, Roland J.
and Moser, Simon
and McCreary, Kathleen M.
and Singh, Simranjeet
and Xu, Jinsong
and Jonker, Berend T.
and Kawakami, Roland K.
and Bostwick, Aaron
and Rotenberg, Eli
and Jozwiak, Chris},
title={Giant spin-splitting and gap renormalization driven by trions in single-layer {WS}$_2$/h-$BN$ heterostructures},
journal={Nature Physics},
year={2018},
month={Apr},
day={01},
volume={14},
number={4},
pages={355-359},
issn={1745-2481},
doi={10.1038/s41567-017-0033-4},
url={https://doi.org/10.1038/s41567-017-0033-4}
}

@Article{Zhang2018,
author={Zhang, Hongyun
and Bao, Changhua
and Jiang, Zeyu
and Zhang, Kenan
and Li, Hao
and Chen, Chaoyu
and Avila, Jos{\'e}
and Wu, Yang
and Duan, Wenhui
and Asensio, Maria C.
and Zhou, Shuyun},
title={Resolving Deep Quantum-Well States in Atomically Thin 2{H}-{M}o{T}e$_2$ Flakes by Nanospot Angle-Resolved Photoemission Spectroscopy},
journal={Nano Letters},
year={2018},
month={Aug},
day={08},
publisher={American Chemical Society},
volume={18},
number={8},
pages={4664-4668},
issn={1530-6984},
doi={10.1021/acs.nanolett.8b00589},
url={https://doi.org/10.1021/acs.nanolett.8b00589}
}

@article{Staab2024,
  title = {Symmetry-enforced Fermi surface degeneracies observed in the purported time-reversal symmetry-broken superconductor {L}a{N}i{G}a$_2$},
  author = {Staab, Matthew and Prater, Robert and Sreedhar, Sudheer and Byland, Journey and Mann, Eliana and Zackaria, Davis and Shi, Yunshu and Bowman, Henry J. and Stephens, Andrew L. and Jung, Myung-Chul and Botana, Antia S. and Pickett, Warren E. and Taufour, Valentin and Vishik, Inna},
  journal = {Phys. Rev. B},
  volume = {110},
  issue = {16},
  pages = {165115},
  numpages = {8},
  year = {2024},
  month = {Oct},
  publisher = {American Physical Society},
  doi = {10.1103/PhysRevB.110.165115},
  url = {https://link.aps.org/doi/10.1103/PhysRevB.110.165115}
}

@ARTICLE{9492063,
  author={Wu, Yinan and Fan, Zhi and Fang, Yongchun and Liu, Cunhuan},
  journal={IEEE/ASME Transactions on Mechatronics}, 
  title={An Intelligent AFM Scanning Strategy Based on Autonomous Exploration}, 
  year={2022},
  volume={27},
  number={3},
  pages={1750-1760},
  doi={10.1109/TMECH.2021.3098223}}

@article{10.1063/5.0172682,
    author = {Kang, Seongseok and Park, Junhong and Lee, Manhee},
    title = {Machine learning-enabled autonomous operation for atomic force microscopes},
    journal = {Review of Scientific Instruments},
    volume = {94},
    number = {12},
    pages = {123704},
    year = {2023},
    month = {12},
    issn = {0034-6748},
    doi = {10.1063/5.0172682},
    url = {https://doi.org/10.1063/5.0172682},
}

@article{doi:10.1021/acs.jpca.3c01685,
author = {Arias, Steven and Zhang, Yunlong and Zahl, Percy and Hollen, Shawna},
title = {Autonomous Molecular Structure Imaging with High-Resolution Atomic Force Microscopy for Molecular Mixture Discovery},
journal = {The Journal of Physical Chemistry A},
volume = {127},
number = {29},
pages = {6116-6122},
year = {2023},
doi = {10.1021/acs.jpca.3c01685},
    note ={PMID: 37462432},
URL = { 
        https://doi.org/10.1021/acs.jpca.3c01685
},
}

@article{LIU2023100704,
title = {Autonomous scanning probe microscopy with hypothesis learning: Exploring the physics of domain switching in ferroelectric materials},
journal = {Patterns},
volume = {4},
number = {3},
pages = {100704},
year = {2023},
issn = {2666-3899},
doi = {https://doi.org/10.1016/j.patter.2023.100704},
url = {https://www.sciencedirect.com/science/article/pii/S2666389923000417},
author = {Yongtao Liu and Anna N. Morozovska and Eugene A. Eliseev and Kyle P. Kelley and Rama Vasudevan and Maxim Ziatdinov and Sergei V. Kalinin}
}

@article{10.1063/5.0082956,
    author = {McDannald, Austin and Frontzek, Matthias and Savici, Andrei T. and Doucet, Mathieu and Rodriguez, Efrain E. and Meuse, Kate and Opsahl-Ong, Jessica and Samarov, Daniel and Takeuchi, Ichiro and Ratcliff, William and Kusne, A. Gilad},
    title = {On-the-fly autonomous control of neutron diffraction via physics-informed Bayesian active learning},
    journal = {Applied Physics Reviews},
    volume = {9},
    number = {2},
    pages = {021408},
    year = {2022},
    month = {04},
    issn = {1931-9401},
    doi = {10.1063/5.0082956},
    url = {https://doi.org/10.1063/5.0082956},
}

@Article{Thomas2022,
author={Thomas, John C.
and Rossi, Antonio
and Smalley, Darian
and Francaviglia, Luca
and Yu, Zhuohang
and Zhang, Tianyi
and Kumari, Shalini
and Robinson, Joshua A.
and Terrones, Mauricio
and Ishigami, Masahiro
and Rotenberg, Eli
and Barnard, Edward S.
and Raja, Archana
and Wong, Ed
and Ogletree, D. Frank
and Noack, Marcus M.
and Weber-Bargioni, Alexander},
title={Autonomous scanning probe microscopy investigations over WS2 and Au{\{}111{\}}},
journal={npj Computational Materials},
year={2022},
month={May},
day={02},
volume={8},
number={1},
pages={99},
issn={2057-3960},
doi={10.1038/s41524-022-00777-9},
url={https://doi.org/10.1038/s41524-022-00777-9}
}

@article{doi:10.1021/acsnano.0c10239,
author = {Vasudevan, Rama K. and Kelley, Kyle P. and Hinkle, Jacob and Funakubo, Hiroshi and Jesse, Stephen and Kalinin, Sergei V. and Ziatdinov, Maxim},
title = {Autonomous Experiments in Scanning Probe Microscopy and Spectroscopy: Choosing Where to Explore Polarization Dynamics in Ferroelectrics},
journal = {ACS Nano},
volume = {15},
number = {7},
pages = {11253-11262},
year = {2021},
doi = {10.1021/acsnano.0c10239},
    note ={PMID: 34228427},
URL = { 
        https://doi.org/10.1021/acsnano.0c10239
},
}

@article{doi:10.1021/acsnano.1c02104,
author = {Kalinin, Sergei V. and Ziatdinov, Maxim and Hinkle, Jacob and Jesse, Stephen and Ghosh, Ayana and Kelley, Kyle P. and Lupini, Andrew R. and Sumpter, Bobby G. and Vasudevan, Rama K.},
title = {Automated and Autonomous Experiments in Electron and Scanning Probe Microscopy},
journal = {ACS Nano},
volume = {15},
number = {8},
pages = {12604-12627},
year = {2021},
doi = {10.1021/acsnano.1c02104},
    note ={PMID: 34269558},
URL = {
        https://doi.org/10.1021/acsnano.1c02104
},
}

@article{10.1063/5.0185362,
    author = {Narasimha, Ganesh and Hus, Saban and Biswas, Arpan and Vasudevan, Rama and Ziatdinov, Maxim},
    title = {Autonomous convergence of STM control parameters using Bayesian optimization},
    journal = {APL Machine Learning},
    volume = {2},
    number = {1},
    pages = {016121},
    year = {2024},
    month = {03},
    issn = {2770-9019},
    doi = {10.1063/5.0185362},
    url = {https://doi.org/10.1063/5.0185362},
}

@article{doi:10.1021/jacs.4c11674,
author = {Zhu, Zhiwen and Yuan, Shaoxuan and Yang, Quan and Jiang, Hao and Zheng, Fengru and Lu, Jiayi and Sun, Qiang},
title = {Autonomous Scanning Tunneling Microscopy Imaging via Deep Learning},
journal = {Journal of the American Chemical Society},
volume = {146},
number = {42},
pages = {29199-29206},
year = {2024},
doi = {10.1021/jacs.4c11674},
    note ={PMID: 39382312},
URL = { 
        https://doi.org/10.1021/jacs.4c11674
},
}

@article{https://doi.org/10.1002/smtd.202301763,
author = {Harris, Sumner B. and Biswas, Arpan and Yun, Seok Joon and Roccapriore, Kevin M. and Rouleau, Christopher M. and Puretzky, Alexander A. and Vasudevan, Rama K. and Geohegan, David B. and Xiao, Kai},
title = {Autonomous Synthesis of Thin Film Materials with Pulsed Laser Deposition Enabled by In Situ Spectroscopy and Automation},
journal = {Small Methods},
volume = {8},
number = {9},
pages = {2301763},
doi = {https://doi.org/10.1002/smtd.202301763},
url = {https://onlinelibrary.wiley.com/doi/abs/10.1002/smtd.202301763},
year = {2024}
}

@article{Ziatdinov_2022,
doi = {10.1088/2632-2153/ac4baa},
url = {https://dx.doi.org/10.1088/2632-2153/ac4baa},
year = {2022},
month = {feb},
publisher = {IOP Publishing},
volume = {3},
number = {1},
pages = {015003},
author = {Ziatdinov, Maxim A and Ghosh, Ayana and Kalinin, Sergei V},
title = {Physics makes the difference: Bayesian optimization and active learning via augmented Gaussian process},
journal = {Machine Learning: Science and Technology}
}

@article{STACH20212702,
title = {Autonomous experimentation systems for materials development: A community perspective},
journal = {Matter},
volume = {4},
number = {9},
pages = {2702-2726},
year = {2021},
issn = {2590-2385},
doi = {https://doi.org/10.1016/j.matt.2021.06.036},
url = {https://www.sciencedirect.com/science/article/pii/S2590238521003064},
author = {Eric Stach and Brian DeCost and A. Gilad Kusne and Jason Hattrick-Simpers and Keith A. Brown and Kristofer G. Reyes and Joshua Schrier and Simon Billinge and Tonio Buonassisi and Ian Foster and Carla P. Gomes and John M. Gregoire and Apurva Mehta and Joseph Montoya and Elsa Olivetti and Chiwoo Park and Eli Rotenberg and Semion K. Saikin and Sylvia Smullin and Valentin Stanev and Benji Maruyama}
}

@article{doi:10.1021/acs.accounts.2c00220,
author = {Seifrid, Martin and Pollice, Robert and Aguilar-Granda, Andr{\'e}s and Morgan Chan, Zamyla and Hotta, Kazuhiro and Ser, Cher Tian and Vestfrid, Jenya and Wu, Tony C. and Aspuru-Guzik, Alán},
title = {Autonomous Chemical Experiments: Challenges and Perspectives on Establishing a Self-Driving Lab},
journal = {Accounts of Chemical Research},
volume = {55},
number = {17},
pages = {2454-2466},
year = {2022},
doi = {10.1021/acs.accounts.2c00220},
    note ={PMID: 35948428},
URL = { 
        https://doi.org/10.1021/acs.accounts.2c00220
},
}

@Article{Merchant2023,
author={Merchant, Amil
and Batzner, Simon
and Schoenholz, Samuel S.
and Aykol, Muratahan
and Cheon, Gowoon
and Cubuk, Ekin Dogus},
title={Scaling deep learning for materials discovery},
journal={Nature},
year={2023},
month={Dec},
day={01},
volume={624},
number={7990},
pages={80-85},
issn={1476-4687},
doi={10.1038/s41586-023-06735-9},
url={https://doi.org/10.1038/s41586-023-06735-9}
}

@Article{Szymanski2023,
author={Szymanski, Nathan J.
and Rendy, Bernardus
and Fei, Yuxing
and Kumar, Rishi E.
and He, Tanjin
and Milsted, David
and McDermott, Matthew J.
and Gallant, Max
and Cubuk, Ekin Dogus
and Merchant, Amil
and Kim, Haegyeom
and Jain, Anubhav
and Bartel, Christopher J.
and Persson, Kristin
and Zeng, Yan
and Ceder, Gerbrand},
title={An autonomous laboratory for the accelerated synthesis of novel materials},
journal={Nature},
year={2023},
month={Dec},
day={01},
volume={624},
number={7990},
pages={86-91},
issn={1476-4687},
doi={10.1038/s41586-023-06734-w},
url={https://doi.org/10.1038/s41586-023-06734-w}
}

@article{KUMAR2025117664,
title = {Towards Gaussian Process for operator learning: An uncertainty aware resolution independent operator learning algorithm for computational mechanics},
journal = {Computer Methods in Applied Mechanics and Engineering},
volume = {435},
pages = {117664},
year = {2025},
issn = {0045-7825},
doi = {https://doi.org/10.1016/j.cma.2024.117664},
url = {https://www.sciencedirect.com/science/article/pii/S0045782524009186},
author = {Sawan Kumar and Rajdip Nayek and Souvik Chakraborty}
}

@article{PhysRevB.107.205119,
  title = {Framework for efficient ab initio electronic structure with Gaussian Process States},
  author = {Rath, Yannic and Booth, George H.},
  journal = {Phys. Rev. B},
  volume = {107},
  issue = {20},
  pages = {205119},
  numpages = {13},
  year = {2023},
  month = {May},
  publisher = {American Physical Society},
  doi = {10.1103/PhysRevB.107.205119},
  url = {https://link.aps.org/doi/10.1103/PhysRevB.107.205119}
}

@Article{Deringer2021,
author={Deringer, Volker L.
and Bart{\'o}k, Albert P.
and Bernstein, Noam
and Wilkins, David M.
and Ceriotti, Michele
and Cs{\'a}nyi, G{\'a}bor},
title={Gaussian Process Regression for Materials and Molecules},
journal={Chemical Reviews},
year={2021},
month={Aug},
day={25},
publisher={American Chemical Society},
volume={121},
number={16},
pages={10073-10141},
issn={0009-2665},
doi={10.1021/acs.chemrev.1c00022},
url={https://doi.org/10.1021/acs.chemrev.1c00022}
}

@article{10.1063/5.0152367,
    author = {Artiukhin, Denis G. and Godtliebsen, Ian H. and Schmitz, Gunnar and Christiansen, Ove},
    title = {Gaussian process regression adaptive density-guided approach: Toward calculations of potential energy surfaces for larger molecules},
    journal = {The Journal of Chemical Physics},
    volume = {159},
    number = {2},
    pages = {024102},
    year = {2023},
    month = {07},
    issn = {0021-9606},
    doi = {10.1063/5.0152367},
    url = {https://doi.org/10.1063/5.0152367},
}

@article{10.1116/6.0004493,
    author = {Kaspar, Tiffany C. and Akers, Sarah and Sprueill, Henry W. and Ter-Petrosyan, Arman H. and Bilbrey, Jenna A. and Hopkins, Derek and Harilal, Ajay and Christudasjustus, Jijo and Gemperline, Patrick and Comes, Ryan B.},
    title = {Machine-learning-enabled on-the-fly analysis of RHEED patterns during thin film deposition by molecular beam epitaxy},
    journal = {Journal of Vacuum Science and Technology A},
    volume = {43},
    number = {3},
    pages = {032702},
    year = {2025},
    month = {03},
    issn = {0734-2101},
    doi = {10.1116/6.0004493},
    url = {https://doi.org/10.1116/6.0004493},
}

@misc{na2025simulationbasedtrainingframeworkmachinelearning,
      title={A simulation-based training framework for machine-learning applications in ARPES}, 
      author={MengXing Na and Chris Zhou and Sydney K. Y. Dufresne and Matteo Michiardi and Andrea Damascelli},
      year={2025},
      eprint={2508.15983},
      archivePrefix={arXiv},
      primaryClass={cond-mat.mtrl-sci},
      url={https://arxiv.org/abs/2508.15983}, 
}

@article{Seah_1979, title={Quantitative electron spectroscopy of surfaces: A standard data base for electron inelastic mean free paths in solids}, volume={1}, ISSN={1096-9918}, url={http://dx.doi.org/10.1002/sia.740010103}, DOI={10.1002/sia.740010103}, number={1}, journal={Surface and Interface Analysis}, publisher={Wiley}, author={Seah, M. P. and Dench, W. A.}, year={1979}, month=feb, pages={2–11} }

@article{Tanuma_1990, title={Electron inelastic mean free paths in solids at low energies}, volume={52}, ISSN={0368-2048}, url={http://dx.doi.org/10.1016/0368-2048(90)85024-4}, DOI={10.1016/0368-2048(90)85024-4}, journal={Journal of Electron Spectroscopy and Related Phenomena}, publisher={Elsevier BV}, author={Tanuma, S. and Powell, C.J. and Penn, D.R.}, year={1990}, month=jan, pages={285–291} }

@article{Rotenberg_2014, title={microARPES and nanoARPES at diffraction-limited light sources: opportunities and performance gains}, volume={21}, ISSN={1600-5775}, url={http://dx.doi.org/10.1107/s1600577514015409}, DOI={10.1107/s1600577514015409}, number={5}, journal={Journal of Synchrotron Radiation}, publisher={International Union of Crystallography (IUCr)}, author={Rotenberg, Eli and Bostwick, Aaron}, year={2014}, month=aug, pages={1048–1056} }

@misc{sreedhar2025mesoscalevariationschemicalelectronic,
      title={Mesoscale variations of chemical and electronic landscape on the surface of Weyl semimetal {C}o$_3${S}n$_2${S}$_2$ visualized by {ARPES} and {XPS}}, 
      author={Sudheer Anand Sreedhar and Matthew Staab and Mingkun Chen and Robert Prater and Zihao Shen and Giuseppina Conti and Ittai Sidilkover and Zhenghong Wu and Eli Rotenberg and Aaron Bostwick and Chris Jozwiak and Hadas Soifer and Slavomir Nemsak and Sergey Y. Savrasov and Vsevolod Ivanov and Valentin Taufour and Inna M. Vishik},
      year={2025},
      eprint={2508.01826},
      archivePrefix={arXiv},
      primaryClass={cond-mat.mtrl-sci},
      url={https://arxiv.org/abs/2508.01826}, 
}

@article{PhysRevB.81.104521,
  title = {Controlling the carrier concentration of the high-temperature superconductor ${\text{Bi}}_{2}{\text{Sr}}_{2}{\text{CaCu}}_{2}{\text{O}}_{8+\ensuremath{\delta}}$ in angle-resolved photoemission spectroscopy experiments},
  author = {Palczewski, A. D. and Kondo, Takeshi and Wen, J. S. and Xu, G. Z. J. and Gu, G. and Kaminski, A.},
  journal = {Phys. Rev. B},
  volume = {81},
  issue = {10},
  pages = {104521},
  numpages = {6},
  year = {2010},
  month = {Mar},
  publisher = {American Physical Society},
  doi = {10.1103/PhysRevB.81.104521},
  url = {https://link.aps.org/doi/10.1103/PhysRevB.81.104521}
}

@article{Biswas_2015, title={Anomalies of a topologically ordered surface}, volume={5}, ISSN={2045-2322}, url={http://dx.doi.org/10.1038/srep10260}, DOI={10.1038/srep10260}, number={1}, journal={Scientific Reports}, publisher={Springer Science and Business Media LLC}, author={Biswas, Deepnarayan and Thakur, Sangeeta and Ali, Khadiza and Balakrishnan, Geetha and Maiti, Kalobaran}, year={2015}, month=jun }

@article{Bagus_2013, title={The interpretation of XPS spectra: Insights into materials properties}, volume={68}, ISSN={0167-5729}, url={http://dx.doi.org/10.1016/j.surfrep.2013.03.001}, DOI={10.1016/j.surfrep.2013.03.001}, number={2}, journal={Surface Science Reports}, publisher={Elsevier BV}, author={Bagus, Paul S. and Ilton, Eugene S. and Nelin, Connie J.}, year={2013}, month=jun, pages={273–304} }

@article{Konig:QSHI2008,
author = {K\"{o}nig ,Markus and Buhmann ,Hartmut and W. Molenkamp ,Laurens and Hughes ,Taylor and Liu ,Chao-Xing and Qi ,Xiao-Liang and Zhang ,Shou-Cheng},
title = {The Quantum Spin Hall Effect: Theory and Experiment},
journal = {Journal of the Physical Society of Japan},
volume = {77},
number = {3},
pages = {031007},
year = {2008},
doi = {10.1143/JPSJ.77.031007},

URL = {

        https://doi.org/10.1143/JPSJ.77.031007



}}

@article{Tokura_2019, title={Magnetic topological insulators}, volume={1}, ISSN={2522-5820}, url={http://dx.doi.org/10.1038/s42254-018-0011-5}, DOI={10.1038/s42254-018-0011-5}, number={2}, journal={Nature Reviews Physics}, publisher={Springer Science and Business Media LLC}, author={Tokura, Yoshinori and Yasuda, Kenji and Tsukazaki, Atsushi}, year={2019}, month=jan, pages={126–143} }
\end{document}


\title{\LARGE \bf Supplementary Materials: Electronic Phase Identification by Photoemission Experiments Using Unsupervised Machine Learning}

\author{Matthew Staab}
	\affiliation{Department of Physics and Astronomy, University of California, Davis, Davis, California $95616$, USA}
\author{Joseph Pandur}
	\affiliation{Department of Physics and Astronomy, University of California, Davis, Davis, California $95616$, USA}
\author{Eli Rotenberg}
	\affiliation{Advanced Light Source, Lawrence Berkeley National Laboratory, Berkeley, CA, USA}
\author{Chris Jozwiak}
	\affiliation{Advanced Light Source, Lawrence Berkeley National Laboratory, Berkeley, CA, USA}
\author{Aaron Bostwick}
	\affiliation{Advanced Light Source, Lawrence Berkeley National Laboratory, Berkeley, CA, USA}
\author{Inna Vishik}
	\affiliation{Department of Physics and Astronomy, University of California, Davis, Davis, California $95616$, USA}

\date{\today}

\maketitle


\section{AARDVARK Software Design}

Adaptive Automation and Real Time Data Visualization for All Research Kit (AARDVARK) is implemented with a communication protocol by passing JSON messages through a ZeroMQ socket. In order to provide as fast of communication as possible, large datasets are saved directly from LabView as a binary file and then read by their filename by the application when needed. When the experiment startup message is received by the Python server, the following two steps are initiated: (1) a new experiment is added to the database to track all decisions, measurements, datasets, and reports, and (2) an `experiment watcher' is initiated in a separate process through Celery to provide new measurement parameters. The initial decision for all experiments consists of a low density random sampling (Poisson disk) of the area. The selected machine learning (ML) pipeline will then take over and create measurement recommendations as fast as they can be calculated. Upon each new recommendation being completed, all unmeasured suggestions are overwritten with the new suggestions and the process is repeated. The recommendation batch size is dynamically tuned such that each new batch will have at least enough suggestions to keep the experiment measuring while the next batch is predicted. This scheme gives balance to the computational cost of suggesting too many measurements while always having enough ready to keep the experiment going. 

The architecture of AARDVARK was specifically designed to accommodate a wide range of modular ML pipelines. The only restrictions of the pipeline are: (1) the first node must accept the $x$ (measurement positions) and $y$ (spectra) that have been collected, and (2) the last node of the pipeline must output a list of suggested measurement positions. This versatility allows for the pipeline discussed later in this work while allowing for future expansion with minimal changes to the codebase.

The source of truth for the experiment is stored in a PostgreSQL database, where experiments are tracked in a method similar to how a user may track their experiment to maximize the ability of users to follow the logic of the autonomous experiment. In AARDVARK, experiments have two main types of messages attached to them: (1) Reports, which may provide some visualization of a model or description of the current status, and (2) Decisions, which contain batches of Measurements that either have been measured and have data attached to them or have not yet been measured and are suggestions for the experiment to perform next. This mimics a user who may \textit{decide} to perform a scan in certain location(s) and \textit{reports} to the group with some visualization of the current status of the experiment.

For the results shown in this work, the ML pipeline used was a combination of dimensional reduction and Gaussian process regression. The dimensional reduction was done using the RAPIDS cuML GPU accelerated version of UMAP \cite{UMAP, UMAP_GPU}. This version of UMAP allows us to reduce approximately 1 million parameter spectra of an ARPES image to a three-dimensional representation in $<1s$ for $\approx 500$ spectra and $<10s$ for $\approx 3000$ spectra. The Gaussian process then fits the three-dimensional representation, effectively predicting the spectra at every possible measurement position. This Gaussian process is implemented using an ExactGP Multitask model from GPytorch \cite{GPYTORCH, MULTITASK_GP} with $100$ training iterations. Using a normalized combination of the standard deviations of the GP predictions, we then suggest a new batch of coordinates to measure next. We suggest the position where the spectrum is least certain as the next measurement, followed by a batch of positions which are generated by sampling the uncertainty map as a probability distribution.

\section{Boundary determination}
For Fig. 6 in the main text, the boundaries were determined from the UMAP plots in Figs. 2 and 5 in the following way. For two pixels with normalized RGB values $(R_1, G_1, B_1)$ and $(R_2, G_2, B_2)$, the color difference $\Delta C$ can be quantified using the 'redmean' formula
\begin{equation} \label{color-diff}
    \Delta C = \sqrt{( 2 + \bar{r}) \Delta R^2 + 4\Delta G^2 + (3-\bar{r}) \Delta B^2}
\end{equation}
where
\begin{equation} \nonumber
    \begin{split}
        & \Delta R = R_1 - R_2\\
        & \Delta G = G_1 - G_2\\
        & \Delta B = B_1 - B_2\\
        & \bar{r} = \frac{1}{2}(R_1+R_2)
    \end{split}
\end{equation}

Each pixel was compared with its closest neighbors and if any pair had a color difference that exceeded a given threshold, then the pixel was determined to represent a boundary. Fig. 6 in the main text shows the resulting boundaries for a sample region of interest.

Table 1 in the main text shows the precision and recall values, with column names defined as follows:
\begin{itemize}
    \item Prec: Precision defined as TP (Prec) / [TP (Prec) + FP].
    \item TP (Prec): True Positives for Precision, or the number of predicted boundary pixels that line up with the ground truth boundary.
    \item FP: False Positives, or the number of predicted boundary pixels that do not line up with the ground truth boundary.
    \item Rec: Recall defined as TP (Rec) / [TP (Rec) + FN]
    \item TP (Rec): True Positives for Recall, or the number of ground truth boundary pixels that line up with the predicted boundary.
    \item FN: False Negatives, or the number of ground truth boundary pixels that do not line up with the predicted boundary.
\end{itemize}

\bibliography{bibliography}